\documentclass[twocolumn]{aastex63}
\usepackage{graphicx}
\usepackage{cancel}
\usepackage{soul}

\newcommand{\teff}{\mbox{$T_{\rm eff}$}}
\newcommand{\logg}{\mbox{$\log g$}}
\newcommand{\feh}{\mbox{$\rm{[Fe/H]}$}}

\usepackage{lineno}
\usepackage{float}
\accepted{02, 2022}
\submitjournal{ApJ}

\begin{document}

\title{Ages of Main-Sequence Turn-Off Stars from the GALAH Survey}

\correspondingauthor{Shaolan Bi, Zhishuai Ge}
\email{E-mail: bisl@bnu.edu.cn, gezhishuai@mail.bnu.edu.cn}
\author{Xunzhou Chen}
\affil{Department of Astronomy, Beijing Normal University, Beijing 100875, China}

\author{Zhishuai Ge}
\affiliation{Beijing Planetarium, Beijing Academy of Science and Technology, Beijing, 100044, China}

\author{Yuqin Chen}
\affiliation{Key Laboratory of Optical Astronomy, National Astronomical Observatories, Chinese Academy of Sciences, A20 Datun Rd, Chaoyang District, Beijing, 100101, China}
\affiliation{School of Astronomy and Space Science, University of Chinese Academy of Sciences, Beijing 100049, China}

\author{Shaolan Bi}
\affiliation{Department of Astronomy, Beijing Normal University, Beijing 100875, China}

\author{Jie Yu}
\affiliation{Max Planck Institute for Solar System Research, Justus-von-Liebig-Weg 3, D-37077 Gottingen, Germany}

\author{Wuming Yang}
\affiliation{Department of Astronomy, Beijing Normal University, Beijing 100875, China}

\author{Jason W. Ferguson}
\affiliation{Department of Physics, Wichita State University, Wichita, KS 67260-0032, USA}

\author{Yaqian Wu}
\affiliation{Key Laboratory of Optical Astronomy, National Astronomical Observatories, Chinese Academy of Sciences, A20 Datun Rd, Chaoyang District, Beijing, 100101, China}

\author{Yaguang Li}
\affiliation{Sydney Institute for Astronomy (SIfA), School of Physics, University of Sydney, NSW 2006, Australia}
\affiliation{Stellar Astrophysics Centre, Department of Physics and Astronomy, Aarhus University, Ny Munkegade 120, 8000 Aarhus C, Denmark}

\begin{abstract}
Main sequence turn-off (MSTO) stars are good tracers of Galactic populations since their ages can be reliably estimated from atmospheric parameters. Based on the GALAH survey, we use the Yale Rotation Evolution Code to determine ages of 2926 MSTO stars with mean age uncertainty $\sim$10\% considering the variation of C and O abundances. Ages of CO-poor stars are systematically affected by $\sim$10\% due to the C and O abundances, globally shifting to $\sim$0.5 Gyr older compared to the results using solar metal-mixture. Of the stars with \mbox{[Fe/H] $\sim$0.3-0.5} or \mbox{[O/Fe]~$\leq$~-0.25}, many have fractional age differences~$\geq$~20\%, and even reach up to 36\%. The age-metallicity relation appears to possibly exist two distinct sequences: a young sequence of stars with age mostly $<$ 7 Gyr, and a relatively older sequence of stars with age mostly $>$ 7 Gyr, overlapping at 5 Gyr $\leq$~age~$\leq$ 7 Gyr. Moreover, the trends of abundances to age ratios show two corresponding sequences, especially in [O/Fe]-age plane. We also find that [Y/Mg] is a good chemical clock in disk populations. The young sequence and the old sequence can not be separated based on chemistry or kinematics, therefore stellar age is an important parameter to distinguish these two sequences in our sample.
\end{abstract}
\keywords{Stars: age --- Stars:  abundances --- Galaxy: disk}
\section{Introduction}
The Galactic evolution history is imprinted in positions, velocities, and chemical abundances of its stars. Since \citet{Gilmore1983} first divide the Galactic disk into a thin disk and a thick disk with different scale height, many works study different populations in the Galactic disk based on chemistry or kinematics \citep[e.g.][]{Adibekyan2012,Silva Aguirre2018}. However, there have also been doubts about whether the thick disk exists \citep{Bovy2012}. It is still an active discussion on how the thin and thick disk components should be defined, and a thorough understanding of the formation and evolution of the Galactic disk requires precise chemical abundances, kinematics, and stellar ages.

The original spatial distribution of stars is changed by the kinematic evolution of the Galaxy, therefore many works suggest that the stellar age rather than kinematics is a better parameter to study different disk populations \citep[e.g.][]{Haywood2013,Bensby2014,Delgado2019}. The relations between abundances and stellar age which can contribute to future models of the Galactic chemical evolution (GCE) are widely studied \citep{Buder2019,Delgado2019,Hayden2020,Sharma2020}. The age structure of different disk populations are shown, providing clues of formation history of the Galactic disk \citep{Haywood2013}. Some abnormal populations such as young $\alpha$-rich stars \citep{Martig2015,Chiappini2015} and old metal-rich stars \citep{Chen2003,Chen2008} are found, indicating the complex evolution history of the Galactic disk. It is of high importance to describe existing populations in Galactic disk combining with reliable age measurements of stars. In recent years, a great effort has been done to derive reliable stellar age \citep[e.g.][]{Nissen2017,Silva Aguirre2018,Delgado2019}. The advent of asteroseismology provides great advance in obtaining stellar age \citep[e.g.][]{Silva Aguirre2018}, but the sample is limited to specific stars such as red giants. Using grid-based stellar evolution model to determine stellar ages is a common and reliable way.

In stellar evolution model, metal-mixture pattern is a crucial part which can affect the opacity. The solar metal-mixture pattern (hereafter solar-mixture) \citep{Grevesse&Sauval1998} and the $\alpha$-enhanced metal-mixture pattern (hereafter $\alpha$-mixture) are widely used in stellar evolution model, e.g. YY isochrones \citep{Yi2001,Yi2003,Kim2002,Demarque2004} and Dartmouth Stellar Evolution Database \citep{Dotter2008}. In the past decades, many observations show that the O abundance shows different behavior from other $\alpha$ elements \citep{Bensby2005,Reddy2006,Nissen2014,Bertran2015,Delgado2019,Amarsi2019,Pavlenko2019,Franchini2021}, and C enhancement also exists \citep{Bensby2005,Reddy2006,Nissen2014}. To study stars with C and O enhancements, the CO-extreme metal-mixture pattern (hereafter CO-mixture), in which the enhancement factors of C and O are added individually and other abundances are consistent with solar-mixture or $\alpha$-mixture, is proposed \citep{Ge2016}. \citet{Ge2016} use CO-mixture to study a sample of halo stars and find that C and O could influence the age determination of stars with C and O enhancements. Based on a sample of disk stars, \citet{Chen2020} find that ages of stars with [O/$\alpha$] $\geq$ 0.2 are obviously younger by $\sim$ 1 Gyr than those by $\alpha$-mixture, and the age difference can affect the [$\alpha$/Fe]-age relation. Furthermore, there are many stars with C and O abundance deficiency, and their age determination should also be affected by C and O abundances. However, due to the lack of high-resolution spectra with precise C and O abundances, the previous samples of \citet{Ge2016} and \citet{Chen2020} are limited in size and mainly comprised of stars with C and O enhancements that were compiled from various sources.

The GALAH DR3 catalogue \citep{buder2021} provides atmospheric parameters and element abundances including precise and homogeneous C and O abundance for 588,571 unique stars that are mainly nearby, enabling us to study a large and self-consistent sample with age, abundance, and kinematics. Furthermore, we can complement our previous studies aiming to understand the influence of C and O enhancements on the stellar age determination by including stars with C and O abundance deficiency.

The MSTO stars are good tracers of Galactic populations and star clusters \citep{Mackey2008,Goudfrooij2009,Yang2013,Wu2017}. The \teff\ of MSTO stars are sensitive to their ages at fixed [Fe/H], therefore their ages can be reliably obtained based on accurate atmospheric parameters. Moreover, the surface chemical abundances of MSTO stars are nearly primordial, essentially without contamination by nuclear reactions in the stellar interior \citep{Nissen2013}. Thus, we select 2926 MSTO stars from GALAH DR3 as sample stars and determine their ages considering the variation of C and O abundances. We show the impact on stellar evolution tracks and age determination by C and O elements. We study abundance to age ratios and the chemical clocks in Galactic disk. With Gaia EDR3 database, we determine and analyze the kinematic properties of sample stars. In Section \ref{sec2}, we show the target selection. In Section \ref{sec3}, we present the stellar evolution model. In Section \ref{sec4}, we describe the results, including chemical and kinematic analyze. In Section \ref{sec5}, we summarise our findings.

\begin{figure}[htbp]
 \centering
 \includegraphics[width=80mm]{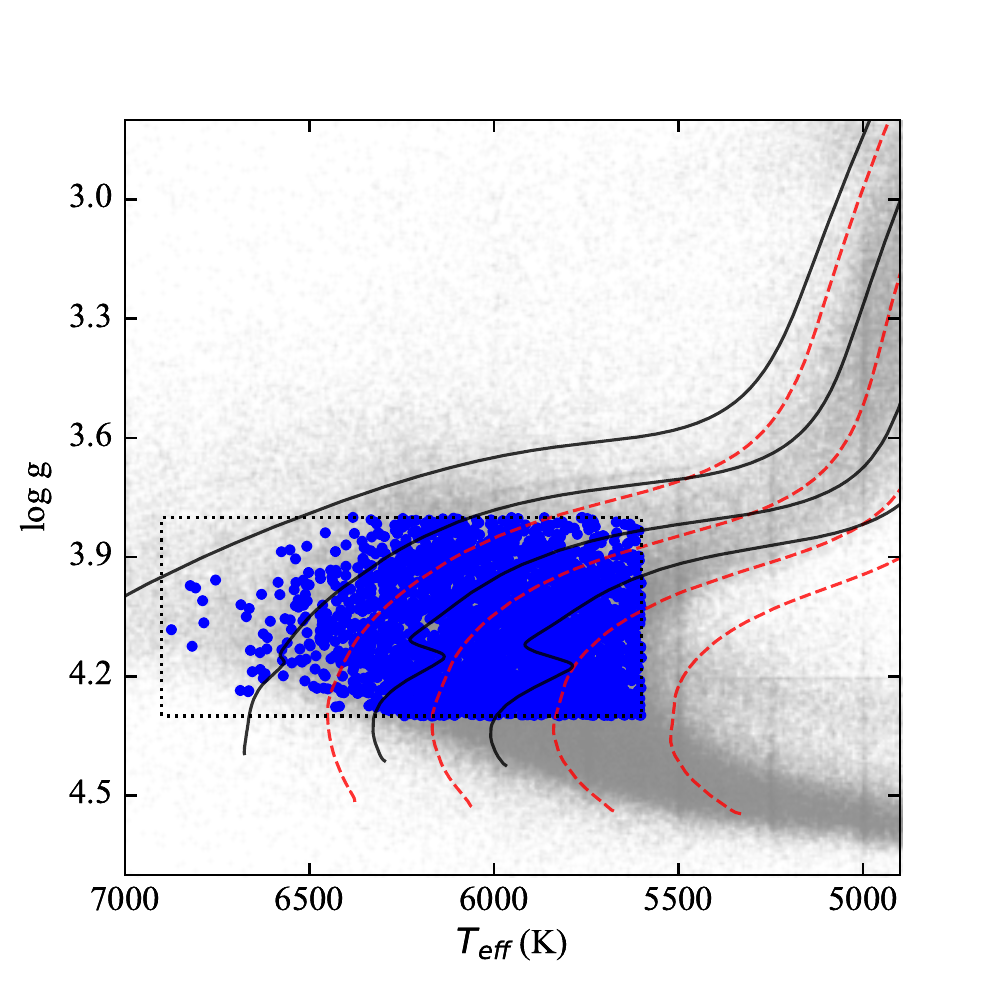}
 \caption{Kiel diagram of the stars from the GALAH DR3 data (gray dots) and the targets used in our work (blue circles). The main-sequence turn-off is delimited by black dotted lines (3.8 $\leq$ \logg\ $\leq$ 4.3, 5600 K $\leq$ \teff\ $\leq$ 6900 K). Red dashed lines show 1.0 $M_\odot$ evolutionary tracks and black solid lines show 1.2 $M_\odot$ evolutionary tracks. At a given mass, metallicities ranges from -0.6 through 0.3 in steps of 0.3 (from left to right).
 }
 \label{fig: fig1}
\end{figure}

\begin{figure*}
 \centering
 \includegraphics[width=19cm]{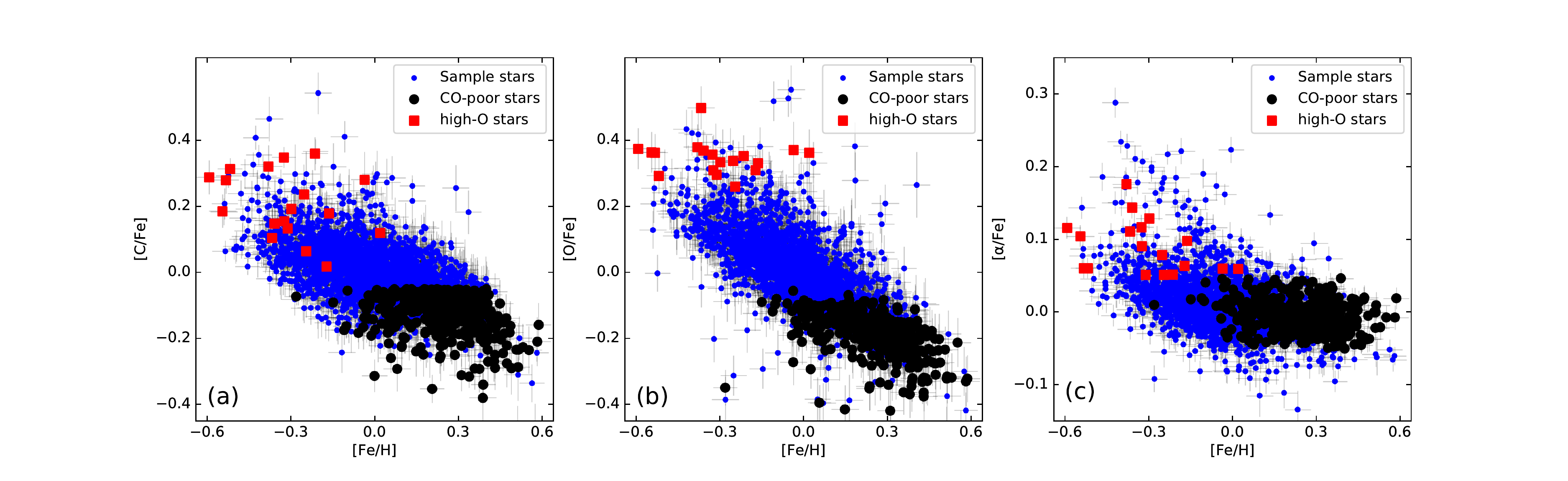}
 \caption{[C/Fe], [O/Fe], and [$\alpha$/Fe] as a function of [Fe/H] for the same stars shown in blue in Figure~\ref{fig: fig1}. Red squares: high-O stars. Blue points: sample stars. Black circles: CO-poor stars.}
 \label{fig2}
\end{figure*}

\begin{table*}
  \tiny
  \centering
  \caption{The observational atmospheric parameters, [C/H], [O/H] and typical errors for temperature, \logg, and [Fe/H] of the sample stars.}
  \label{tab1}
  \renewcommand\tabcolsep{4.0pt} %
  \begin{tabular}{ccccccc}
  \hline
  Star &\teff & \logg & $[\rm{Fe/H}]$ & $[\rm{C/H}]$ & $[\rm{O/H}]$&$[\rm{\alpha/Fe}]$ \\

   ID & (K) & (dex) & (dex)  & (dex) & (dex) & (dex)   \\																																							
  \hline
11285775-0440070	&	6176 	$ \pm $	72 	&	4.16 	$ \pm $	0.06 	&	-0.15 	$ \pm $	0.05 	&	-0.24	&	-0.24&	0.02  \\
17215393+0618245	&	6000 	$ \pm $	72 	&	4.29 	$ \pm $	0.07 	&	0.21 	$ \pm $	0.05 	&	-0.06 	&   0.13 &   0.02 	\\

  $\vdots$&$\vdots$&$\vdots$&$\vdots$&$\vdots$&$\vdots$&$\vdots$\\
  \hline
  \end{tabular}
  {\\Note. This table is available in its entirety in machine-readable form.}
\end{table*}

\section{Target Selection}\label{sec2}
\subsection{Main-Sequence Turn-Off stars}\label{sec2.1}
We select our targets from the GALAH DR3 catalogue \citep{buder2021}. This catalogue provides \teff, \logg, [Fe/H], and up to 30 element abundances for 588,571 unique stars. Along with the other elements, the C and O abundances are derived through non-LTE computation. Firstly, we only keep stars without bad flags in \texttt{flag\_sp, flag\_fe\_h, flag\_alpha\_fe, flag\_c\_e, flag\_o\_fe, flag\_mg\_fe, flag\_si\_fe, flag\_ca\_fe, flag\_ti\_fe}. For precision, we then prune our sample by demanding the formal uncertainties of [Fe/H], [C/Fe], and [O/Fe] less than 0.05, 0.1, and 0.1, respectively. According to \citet{Bonaca2020}, we apply their criteria 3.8~$\leq$~\logg~$\leq$~4.3 to select MSTO stars. Finally, similarly as \citet{Buder2019}, we remove hot stars with \teff~$\geq$~6900 K and evolved stars with \logg~$\leq$~3.8, \teff~$\leq$~5600 K, leaving us with 3025 targets. Figure~\ref{fig: fig1} shows the distribution of the selected MSTO stars in the Kiel diagram. Table \ref{tab1} shows the atmospheric parameters, [C/H], [O/H] and typical errors for temperature, \logg, and [Fe/H] of the sample stars.

\subsection{Stars with extreme [C/Fe] and [O/Fe]}

Our sample contains 18 high-O stars, defined as [O/$\alpha$]~$\geq$~0.2 and [$\alpha$/Fe]~ $\geq$~0.05, according to \citet{Chen2020}. Our sample also incorporate stars with negative values of [C/Fe] and [O/Fe], which are hereafter referred to as \mbox{CO-poor} stars. We use the following criteria to define the CO-poor stars:
\begin{enumerate}
\item[1.] [C/Fe] $\leq$ -0.05 and [O/Fe] $\leq$ -0.05
\item[2.] [O/$\alpha$] $\leq$ -0.1
\item[3.] -0.05 $\leq$ [$\alpha$/Fe] $\leq$ 0.05
\end{enumerate}
These criteria ensure that the CO-poor stars have truly negative values of [C/Fe] and [O/Fe], and their [O/Fe] differs with [$\alpha$/Fe]. We keep their [$\alpha$/Fe]$\sim$0, consistent with the solar-mixture that will be used as a reference for comparison in the following section. With the above criteria, we select 439 CO-poor stars. This sample of 18 high-O stars and 439 \mbox{CO-poor} stars are regarded as stars with extreme [C/Fe] and [O/Fe], enabling our age study into the impact by the variation of C and O abundances on the age determination of stars.

Figure~\ref{fig2} shows relations of [C/Fe], [O/Fe], and [$\alpha$/Fe] versus [Fe/H]. The trends of the [X/Fe] vs. [Fe/H] relations can be interpreted as the result of a chemical enrichment regulated by the timescales of the polluters. Elements such as C, N and
Fe are mainly produced by long living stars (AGB and type \uppercase\expandafter{\romannumeral1}a SNe), while others, such as $\alpha$-elements (O, Mg, Si, Ca) are produced by short living stars (thorugh winds and CCSNe). In particular, Fe is mainly produced by type \uppercase\expandafter{\romannumeral1}a SNe and only a small fraction is ejected by CCSNe. type \uppercase\expandafter{\romannumeral1}a SNe explode on longer timescales than CCSNe and therefore ratios such as [$\alpha$/Fe] can be used as cosmic clocks. The positive values of the [$\alpha$/Fe] ratios at low [Fe/H] is then due to the CCSNe which restore the $\alpha$-elements on short timescales. When type \uppercase\expandafter{\romannumeral1}a SNe, originating from CO white dwarfs, start restoring the bulk of Fe, then the [$\alpha$/Fe] ratios start decreasing. Observations can therefore set constraints on the origin of the different elements, and many works chemically selected high-$\alpha$ and low-$\alpha$ populations in [$\alpha$/Fe]-[Fe/H] plane \citep[e.g.][]{Adibekyan2012,Silva Aguirre2018}. In our sample, the [Fe/H] of sample stars ranges from -0.6 to 0.6, and shows continuous distribution, indicating no clear separation based on chemistry. Most high-O stars have [C/Fe]$\sim$0 to 0.4, [O/Fe]$\sim$0.2 to 0.4, and CO-poor stars have [C/Fe]$\sim$-0.3 to -0.1, [O/Fe]$\sim$-0.4 to -0.1. Almost all the CO-poor stars have [Fe/H]~$\geq$~0. Having a negative value of [O/Fe] at [Fe/H]~$\geq$~0 is typically interpreted as due to the occurrence of type \uppercase\expandafter{\romannumeral1}a SNe that enrich in Fe.

\begin{table*}
  \footnotesize
  \centering
  \renewcommand\tabcolsep{4.0pt} %
  \caption{Metal mixtures for GS98 solar-mixture, $\alpha$-enhanced mixutre ([$\alpha$/Fe] = 0.2) and CO-mixture ([C/Fe]=-0.1, [O/Fe]=-0.2, [$\alpha$/Fe]= 0 in CO-poor case, and [C/Fe]=0.2, [O/Fe]=0.4, [$\alpha$/Fe]= 0.2 in CO-rich case, respectively). The $\alpha$ elements are O, Ne, Mg, Si, S, Ca and Ti, excluding Ar.}
  \label{tab2}
  \begin{tabular}{ccccccc}
  \hline
   Element & $\log N_\odot$ & $\log N_{\alpha}$  & $\log N_{CO}$ (CO-poor) & $\log N_{CO}$ (CO-rich)  \\
    \hline
    \textbf{C}  & 8.52 &8.52&8.52-0.1 &8.52+0.2 \\								
    N  & 7.92 &7.92 &7.92&7.92\\								
    \textbf{O} &  8.83 &8.83+0.2&8.83-0.2&8.83+0.4 \\								
    F  & 4.56  &4.56 &4.56&4.56\\								
    \textbf{Ne}  & 8.08 & 8.08+0.2&8.08&8.08+0.2 \\								
    Na  & 6.33  &6.33 &6.33&6.33\\								
    \textbf{Mg}  & 7.58 &7.58+0.2& 7.58&7.58+0.2  \\								
    Al & 6.47 &6.47&6.47&6.47 \\								
    \textbf{Si}  & 7.55 & 7.55+0.2 & 7.55 & 7.55+0.2\\								
    P  & 5.45 & 5.45& 5.45& 5.45\\								
    \textbf{S}  & 7.33 &7.33+0.2& 7.33&7.33+0.2 \\								
    Cl  & 5.50 & 5.50 & 5.50& 5.50\\								
    Ar  & 6.40 & 6.40 & 6.40& 6.40 \\								
    K  & 5.12 & 5.12 & 5.12& 5.12 \\								
    \textbf{Ca}  & 6.36 & 6.36+0.2& 6.36& 6.36+0.2 \\								
    Sc  & 3.17 & 3.17& 3.17& 3.17\\								
    \textbf{Ti} & 5.02 & 5.02+0.2 & 5.02& 5.02+0.2\\								
    V & 4.00 &4.00&4.00&4.00 \\								
    Cr & 5.67 & 5.67& 5.67& 5.67\\								
    Mn  & 5.39 & 5.39& 5.39& 5.39 \\								
    Fe  & 7.50 & 7.50& 7.50& 7.50 \\								
    Co& 4.92 &  4.92&  4.92&  4.92\\								
    Ni  & 6.25 &  6.25&  6.25&  6.25\\								
  \hline
  \end{tabular}
\end{table*}

\begin{table*}
  \footnotesize
  \centering
  \caption{Grid of evolutionary models with various metal-mixture patterns. The mass step is 0.02 $M_\odot$ and Z step is 0.0010.}
  \label{tab3}
  \begin{tabular}{ccccccc}
  \hline
   Metal-mixture&$[\rm{C/Fe}]$ & $[\rm{O/Fe}]$ & $[\rm{\alpha/Fe}]$ &Mass range &Z range \\
   &(dex) & (dex) & (dex) & $M_\odot$ & heavy element abundance \\	
    \hline
   CO-mixture&-0.1&-0.1&0&0.86$\sim$1.54&0.0110$\sim$0.0460	\\		
   &-0.1&-0.2&0&	0.86$\sim$1.44\ (1.46$\sim$1.54)&0.0110$\sim$0.0420\ (0.0140$\sim$0.0340)	\\			
   &-0.1&-0.3&0&	0.86$\sim$1.44&0.0110$\sim$0.0430		\\
   &-0.2&-0.2&0&	0.86$\sim$1.54&0.0120$\sim$0.0420	\\				 &-0.2&-0.3&0&	0.86$\sim$1.44\ (1.46$\sim$1.54)&0.0120$\sim$0.0450\ (0.0180$\sim$0.0380)									\\		
   &0&0.3&0.1&	0.86$\sim$1.34&0.0060$\sim$0.0310		\\		
   &0.2&0.4&0.2&	0.96$\sim$1.34&0.0060$\sim$0.0310		\\		
   \hline
   solar-mixture&0&0&0&	0.86$\sim$1.54&0.0060$\sim$0.0500			\\	
   \hline
  $\alpha$-mixture &0&0.1&0.1&	0.86$\sim$1.54&0.0100$\sim$0.0360			\\		
   &0&0.2&0.2&	0.76$\sim$1.44\ (1.46$\sim$1.54)&0.0060$\sim$0.0250\ (0.0180$\sim$0.0210)								\\		
  \hline
  \end{tabular}
\end{table*}

\begin{table}
  \footnotesize
  \centering
  \caption{Parameters of 16 models at the end of the main Sequence at fixed Z. Z represents the heavy element abundance.}
  \label{tab4}
  \renewcommand\tabcolsep{1.5pt} %
  \begin{tabular}{ccccccccc}
  \hline
  Model &Mass&  Z &[Fe/H]& [C/Fe] & [O/Fe] &\teff &$\log(L/L_\odot) $  &Age  \\

    &$M_\odot$&   & (dex) & (dex) & (dex)   & (K)&  & (Gyr) \\
  \hline
  1 &1.2& 0.050 &0.53 & 0    & 0     &5641 & 0.32 & 5.65\\
  2 &1.2& 0.050 &0.63 & -0.1 & -0.2  &5522 & 0.25 & 6.73\\
  3 &1.2& 0.050 &0.66 & -0.1 & -0.3  &5483 & 0.23 & 7.09\\
  4 &1.2& 0.050 &0.68 & -0.2 & -0.3  &5463 & 0.22 & 7.30\\
    \hline

  5 &1.2& 0.010&-0.23 & 0    & 0     &6481 & 0.61 & 3.34\\
  6 &1.2& 0.010&-0.13 & -0.1 & -0.2  &6396 & 0.53 & 3.46\\
  7 &1.2& 0.010&-0.09 & -0.1 & -0.3  &6363 & 0.52 & 3.63\\
  8 &1.2& 0.010&-0.08 & -0.2 & -0.3  &6346 & 0.51 & 3.73\\
    \hline

  9 &1.2& 0.005&-0.54 & 0 & 0  &7083 & 0.68 & 2.64\\
  10 &1.2& 0.005&-0.44 & -0.1 & -0.2 &6873 & 0.66 & 2.88\\
  11 &1.2& 0.005&-0.40 & -0.1 & -0.3 &6825 & 0.65 & 2.97\\
  12 &1.2& 0.005&-0.39 & -0.2 & -0.3  &6799 & 0.65 & 3.03\\
    \hline

  13 &1.2& 0.001&-1.24 & 0 & 0  &8542 & 0.86 & 2.33\\
  14 &1.2& 0.001&-1.14 & -0.1 & -0.2  &8461 & 0.86 & 2.41\\
  15 &1.2& 0.001&-1.11 & -0.1 & -0.3  &8434 & 0.86 & 2.44\\
  16 &1.2& 0.001&-1.09 & -0.2 & -0.3  &8419 & 0.86 & 2.46\\		
  \hline
  \end{tabular}
\end{table}

\begin{table}
  \footnotesize
  \centering
  \caption{Parameters of 16 Models at the End of the Main Sequence at fixed [Fe/H]. Z represents the heavy element abundance.}
  \label{tab5}
  \renewcommand\tabcolsep{1.5pt} %
  \begin{tabular}{ccccccccc}
  \hline
  Model &Mass&  Z &[Fe/H]& [C/Fe] & [O/Fe] &\teff &$\log(L/L_\odot) $  &Age  \\
    &$M_\odot$   &&  (dex)    & (dex) & (dex) &(K)&  & (Gyr) \\	
    \hline
  17 &1.2&0.0315& 0.3 & 0 & 0 & 5857 & 0.39 & 5.22\\
  18 &1.2& 0.0255& 0.3 & -0.1 & -0.2 & 5883 & 0.37 & 5.49\\
  19 &1.2& 0.0237& 0.3 & -0.1 & -0.3 & 5893 & 0.32 & 5.02\\
  20 &1.2& 0.0229& 0.3 & -0.2 & -0.3 & 5887 & 0.32 & 5.15\\
    \hline
  21 &1.2&0.0166& 0 & 0 & 0  &6202 & 0.52& 4.17\\
  22 &1.2& 0.0133& 0 & -0.1 & -0.2 &6250 &0.46 & 3.85\\
  23 & 1.2& 0.0123&0 & -0.1 & -0.3 &6258 & 0.47 & 3.93\\
  24 & 1.2& 0.0119&0 & -0.2 & -0.3 &6255 & 0.47 & 4.00\\
    \hline

  25 &1.2&0.0054& -0.5 & 0 & 0 &6968 & 0.67 & 2.66\\
  26 &1.2&0.0043& -0.5 & -0.1 & -0.2 & 7036 & 0.69 & 2.79\\
  27 &1.2&0.0040& -0.5 & -0.1 & -0.3 & 7096 & 0.69 & 2.84\\
  28 &1.2&0.0039& -0.5 & -0.2 & -0.3 &7090 & 0.70 & 2.89\\
    \hline

  29 &1.2&0.0017& -1.0 & 0 & 0 &8121 & 0.11 & 2.35\\
  30 &1.2&0.0014& -1.0 & -0.1 & -0.2 &8175 & 0.11 & 2.44\\
  31 & 1.2&0.0013&-1.0 & -0.1 & -0.3  &8208 & 0.11 & 2.47\\
  32 &1.2&0.0012& -1.0 & -0.2 & -0.3  &8262 & 0.11 & 2.48\\
  \hline
  \end{tabular}
\end{table}

\section{Stellar models}\label{sec3}

\begin{figure*}[htbp]
\centering
\includegraphics[width=0.8\textwidth]{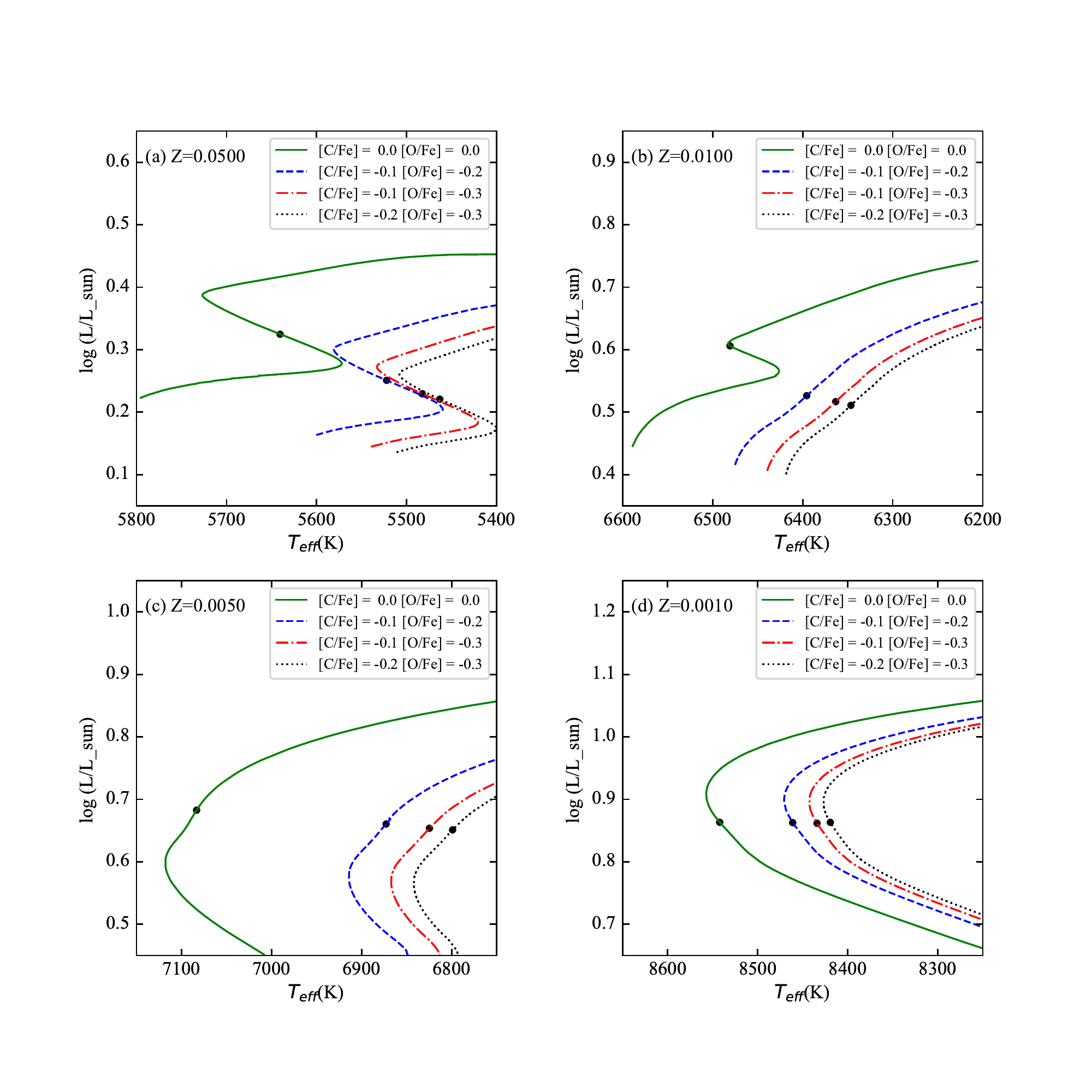}
\caption{Solar-metallicity evolutionary tracks of $M$ = 1.2 M$_\odot$ with multiple [C/Fe] and [O/Fe] combinations at four Z values as indicated. The black point on each track represents the model at the end of the main sequence (X~$=$~0.01).}
\label{fig3}
\end{figure*}

\subsection{Input physics}

We use the Yale Rotation Evolution Code (YREC, \citep{Guenther1992}) in its non-rotation configuration to compute a grid of stellar evolutionary tracks. The helium abundance is calibrated against standard solar models, and thus Y = 0.248 + 1.3324Z \citep{Spergel2007}. The mixing-length parameter $\alpha_{l}$ is fixed to 1.75. Our stellar models are based on the 2005 update of the OPAL EOS tables \citep{Rogers2002}, OPAL opacity tables \citep{Iglesias1996} with GS98 mixture \citep{Grevesse&Sauval1998} at high temperature, and the Ferguson opacity tables \citep{Ferguson2005} at low temperature. We use solar-mixture to determine ages of stars without $\alpha$ enhancements, and $\alpha$-mixture to determine ages of $\alpha$-enhanced stars.

For the high-O and CO-poor stars, we construct CO-mixture to redetermine their ages. Assuming [Fe/H] = 0, for GS98 scaled-solar metal mixture, \textbf{the metallicity Z (in mass fraction) is 0.0166, and} the elements have the same proportion as solar metal elements:
\begin{equation}\label{eq1}
[M/Fe]=[M/Fe]_{\odot}=0
\end{equation}
where $M$ corresponds to the metal element, and the enhancement of a single element could be considered as:
\begin{equation}\label{eq2}
[M/Fe]=\log (\frac{N_M}{N_{Fe}})_{star}-\log (\frac{N_M}{N_{Fe}})_\odot
\end{equation}
where $N$ stands for the number of the particles in a unit volume (e.g. the abundance by number), and $\log N_H=12$. From this relation, we consider $[M/Fe]$ as the enhancement of a metal element to the solar mixture, and the value of [M/Fe] can be calculated from the observed element abundances:
\begin{equation}\label{eq3}
[M/Fe] = [M/H]-[Fe/H]
\end{equation}
We construct CO-mixture by adding enhancement factors to the solar $\log N_i$ values with stable $\log N_{Fe}$ ($N_i$ represents the volume density of the element $i$, in our case they are C, O, and $\alpha$-elements) in the same way as \citet{Ge2015}. For example, Table \ref{tab2} lists metal mixtures for GS98 solar-mixture, $\alpha$-enhanced mixture ([$\alpha$/Fe] = 0.2) and CO-mixture ([C/Fe]=-0.1, [O/Fe]=-0.2, [$\alpha$/Fe]= 0 in CO-poor case, and [C/Fe]=0.2, [O/Fe]=0.4, [$\alpha$/Fe]= 0.2 in CO-rich case, respectively). The OPAL high-temperature opacity tables are constructed online \footnote{http://opalopacity.llnl.gov/new.html} with 3.75~$\leq$~log~$T$~$\leq$~8.7. The low-temperature opacity tables are reconstructed with 2.7~$\leq$~log~$T$~$\leq$~4.5 according to the CO-mixture in a similar way as \citet{Ferguson2005}. The metal-mixture patterns used in this work and the detailed parameters of grid computation are listed in Table \ref{tab3}.

\begin{figure*}[htbp]
\centering
\includegraphics[width=0.8\textwidth]{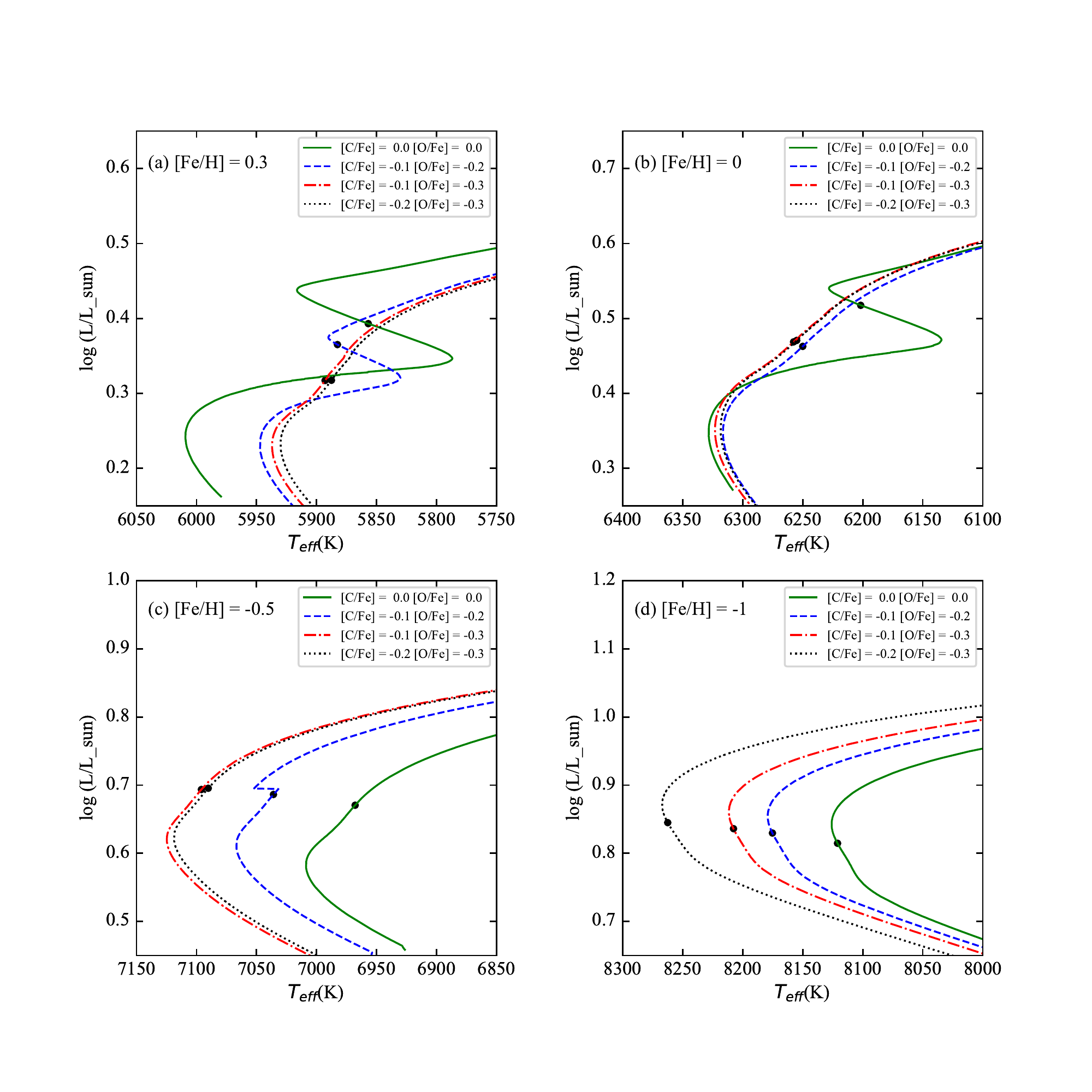}
\caption{Similar as Figure \ref{fig3} but now at four [Fe/H] values rather than Z.}
\label{fig4}
\end{figure*}

\subsection{Effects on Evolutionary Tracks by Variation of C and O Abundances}
Figure \ref{fig3} shows evolutionary tracks with four combinations of [C/Fe] and [O/Fe] for a given Z in each panel. We find that the tracks with lower [C/Fe] and [O/Fe] globally shift to lower \teff\ at a given Z. For each track, we choose one model at the end of the main sequence (X~$\leq$~0.01, black dots in Figure~\ref{fig3}). Table \ref{tab4} lists the parameters of these models (1$\sim$16). From model 1 to 4, the luminosities decrease with decreasing [C/Fe] and [O/Fe] and the ages increase with decreasing [C/Fe] and [O/Fe]. At fixed Z, the variation of [C/Fe] and [O/Fe] would influence opacity, which could influence the energy transfer efficiency and the thermal structure, therefore the lifetime of main-sequence phase is changed.

\begin{figure*}[ht]
\centering
\includegraphics[width=0.9\textwidth]{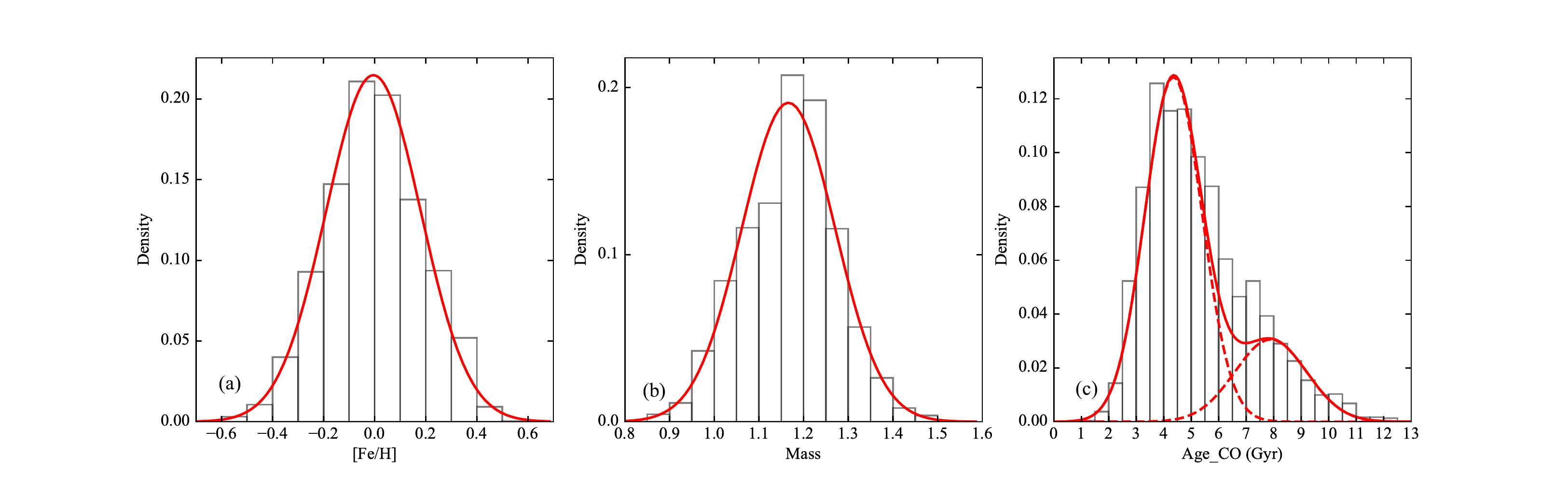}
\caption{Probability density distributions (bars) and their optimal Gaussian fits (red lines) of [Fe/H], mass, and age of our sample stars.}
\label{fig5}
\end{figure*}

\begin{figure}
\centering
\includegraphics[width=0.45\textwidth]{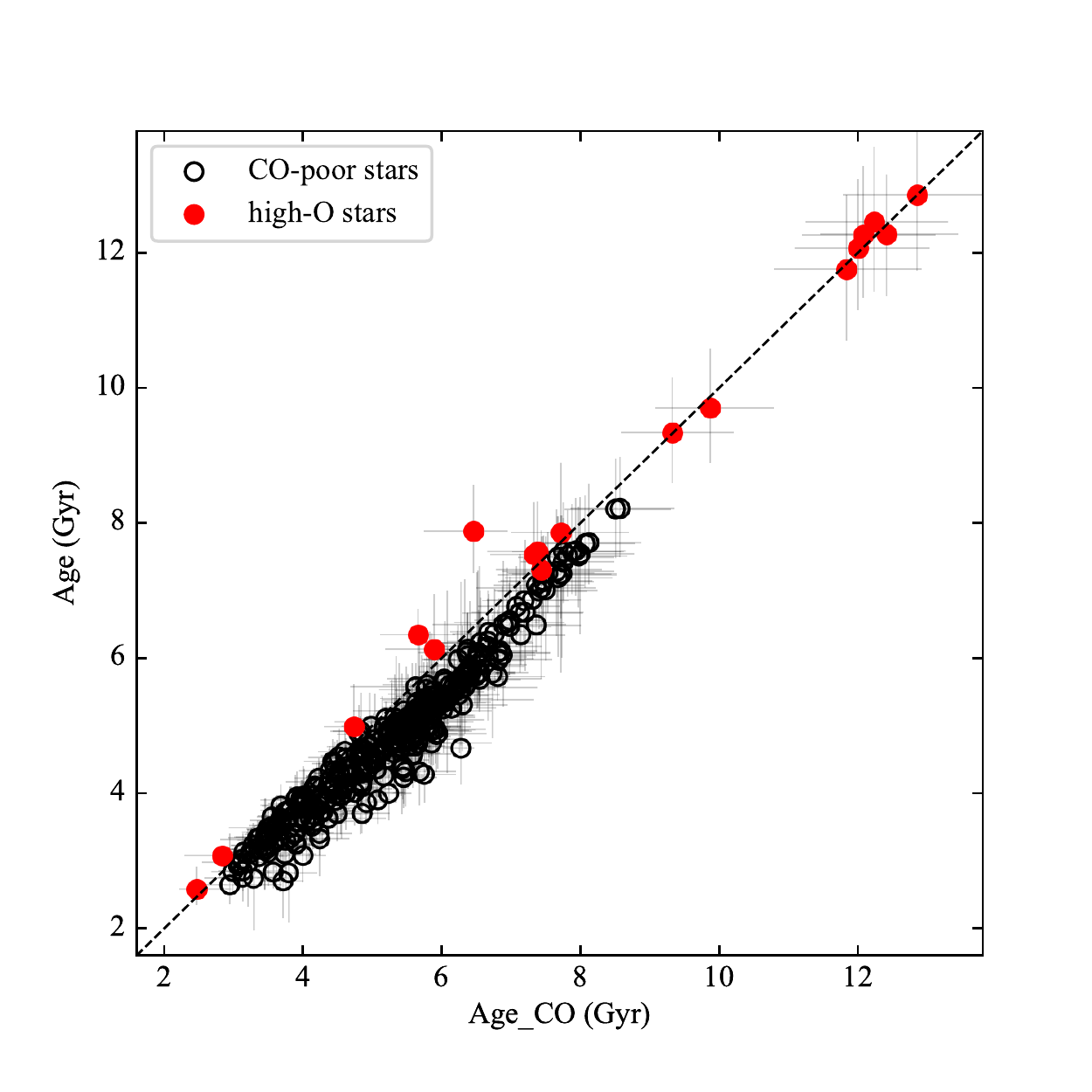}
\caption{Comparison of ages determined by considering solar-scaled and non-solar-scaled C and O mixtures for CO-poor (open black circles) and high-O (solid red circles) stars. See the text for the definitions of the two groups of stars.}
\label{fig6}
\end{figure}

\begin{figure}
\centering
\includegraphics[width=0.45\textwidth]{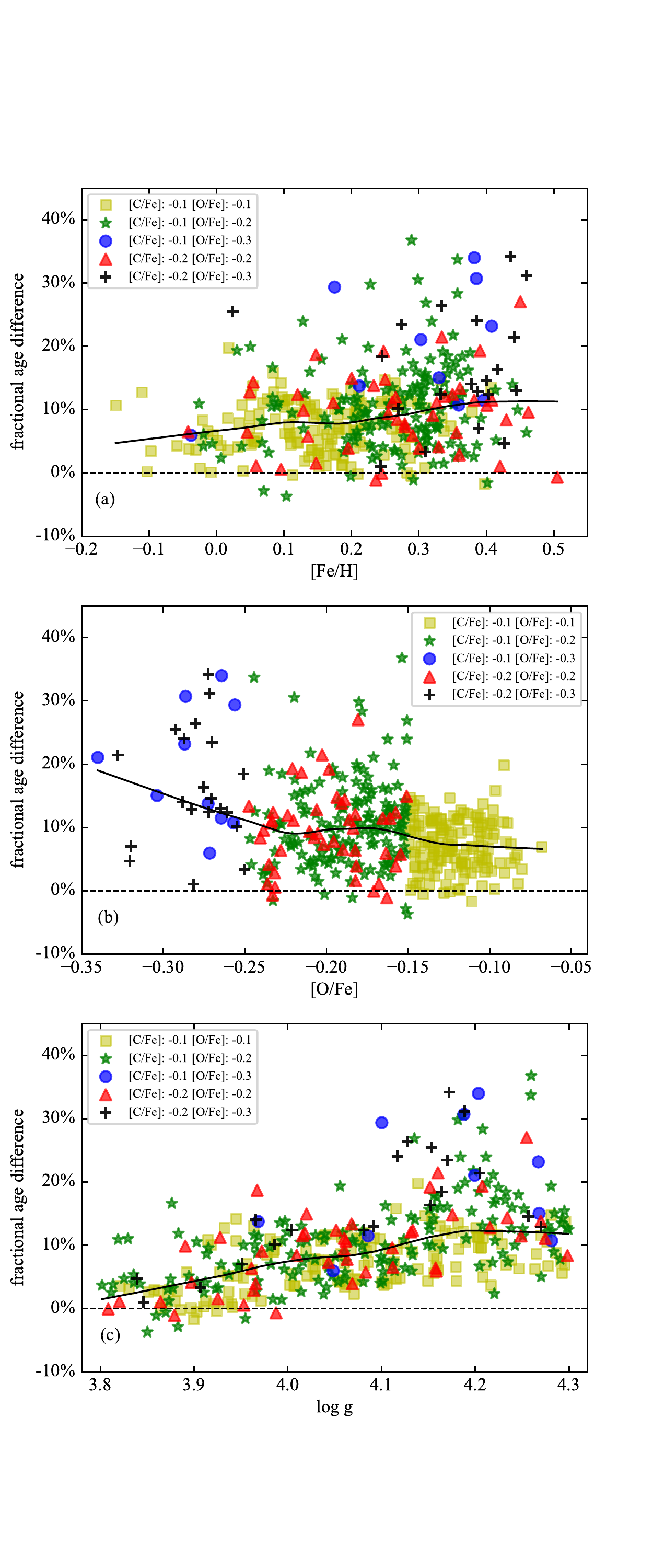}
\caption{Fractional age difference in the sense of $\frac{Age_{CO}~-~Age}{Age}$ as a function of [Fe/H], [O/Fe], and \logg\ (from top to bottom). Red solid lines represent the fitting result by local non-parametric regression. Ages are calculated either with CO-mixtures (see legends) or canonical solar-mixture.}
\label{fig7}
\end{figure}

Figure \ref{fig4} shows tracks of different [C/Fe] and [O/Fe] at a given [Fe/H] in each panel. We find that tracks of [C/Fe] = 0 and [O/Fe] = 0 globally have the lowest temperatures at [Fe/H] = - 1, and gradually become irregular at [Fe/H] = 0.3. We also select models at the end of the main sequence (black dots in Figure~\ref{fig4}) to analyze. Table \ref{tab5} lists the parameters of these models (17$\sim$32). From model 17 to 20, the Z decrease with decreasing [C/Fe] and [O/Fe] for a given [Fe/H], but the lifetime of the main-sequence is not always longer with decreasing [C/Fe] and [O/Fe]. At fixed [Fe/H], the variation of C and O abundances would change the heavy metal abundance Z, resulting in changes of the hydrogen abundance X and the helium abundance Y.

\subsection{Fundamental parameter estimation}
In order to obtain fundamental parameters including stellar age, we use a Bayesian scheme which is similar as \citet{Kallinger2010}, and \citet{Basu2010} to find the most probable stellar models from evolutionary tracks. Based on a set of observed constrains $\o$ (in our case, they are \teff, \logg, and [Fe/H]), we define the likelihood that matches the observed constrains as:
\begin{equation}\label{eq4}
L=\frac{1}{\sqrt{2\pi}\sigma}exp{(\frac{-\chi^2}{2})},
\end{equation}
where
\begin{equation}\label{eq5}
\chi^2=(\frac{\o_{obs}-\o_{model}}{\sigma})^2.
\end{equation}
Here $\sigma$ is the error of the observation $\o_{obs}$.
According to Bayes' theorem, the posterior probability of model $M_{i}$ given data $D$ is computed via:
\begin{equation}\label{eq6}
p(M_{i}|D,I)=\frac{p(M_{i}|I)p(D|M_{i},I)}{p(D|I)}.
\end{equation}
We assume a uniform prior $p(M_{i}|I)\ =\ \frac{1}{N_m}$, where $N_m$ is the total number of computed models. Our likelihood function is defined as:
\begin{equation}\label{eq7}
p(D|M_{i},I)=L(\teff,\logg, \feh)=L_{\teff}L_{\logg}L_{\feh}.
\end{equation}
Since $p(D|I)$ is just a normalization factor and $p(M_{i}|I)$ is constant, we have:
\begin{equation}\label{eq8}
p(M_{i}|D,I) \propto p(D|M_{i},I).
\end{equation}
Thus, maximizing the likelihood function yields the most probable model. We estimate optimal parameters (\teff, \logg, \feh, age, etc.) and their errors by calculating the 16th , 50th, and 84th percentiles of individual marginal posterior distributions.

\section{Result}\label{sec4}
\subsection{Stellar Ages}\label{sa}

We obtain ages of 2926 MSTO stars with mean age uncertainty of $\sim$10\%, including 18 high-O stars and 384 CO-poor stars. Stars with age uncertainty $\geq$20\% have been removed. The fundamental parameters of sample stars are listed in Table \ref{tab6}. Figure \ref{fig5} shows the distribution of [Fe/H], mass, and age of the total sample stars. Our sample stars mainly belong to F and G type stars, having a mean [Fe/H]$\sim$0. The age distribution are well fitted by two gaussian profiles, indicating 2 different groups in our sample: a young group with mean age $\sim$4.5 Gyr, and an relatively old group with mean age $\sim$ 8 Gyr.

Figure \ref{fig6} shows a comparison between ages determined with CO-mixture and solar-mixture or $\alpha$-mixture of high-O and CO-poor stars. Ages calculated with CO-mixture are systematically older than ages with solar-mixture or $\alpha$-mixture by $\sim$0.5 Gyr ($Age_{CO}$~$\geq$~$Age$). Of the 18 high-O stars, 13 are younger after considering CO-mixture, which is consistent with \citet{Chen2020}. Therefore, the C and O abundances can systematically influence the age determination of both CO-poor stars and high-O stars.

Figure \ref{fig7} shows how the fractional differences of age vary with [Fe/H], [O/Fe] and \logg\ for CO-poor stars. We perform local non-parametric regression fitting (LOESS model) for our sample in each panel (black solid lines, the same below). The mean fractional age difference is $\sim$10\%. Figure \ref{fig7}a and \ref{fig7}b show the fractional difference increases with increasing [Fe/H] and decreases with increasing [O/Fe], respectively. Of the stars with \mbox{[Fe/H] $\sim$0.3-0.5} or \mbox{[O/Fe]~$\leq$~-0.25}, many have fractional age differences~$\geq$~20\%, and even reach up to 36\%. Figure~\ref{fig7}c shows that the fractional age difference increases with \logg. Therefore, the impact of C and O abundances on stellar evolution (age) is related to [Fe/H], [O/Fe] and \logg.

\subsection{Chemical Abundance Trends with Age}

\begin{figure}
\centering
\includegraphics[width=0.5\textwidth]{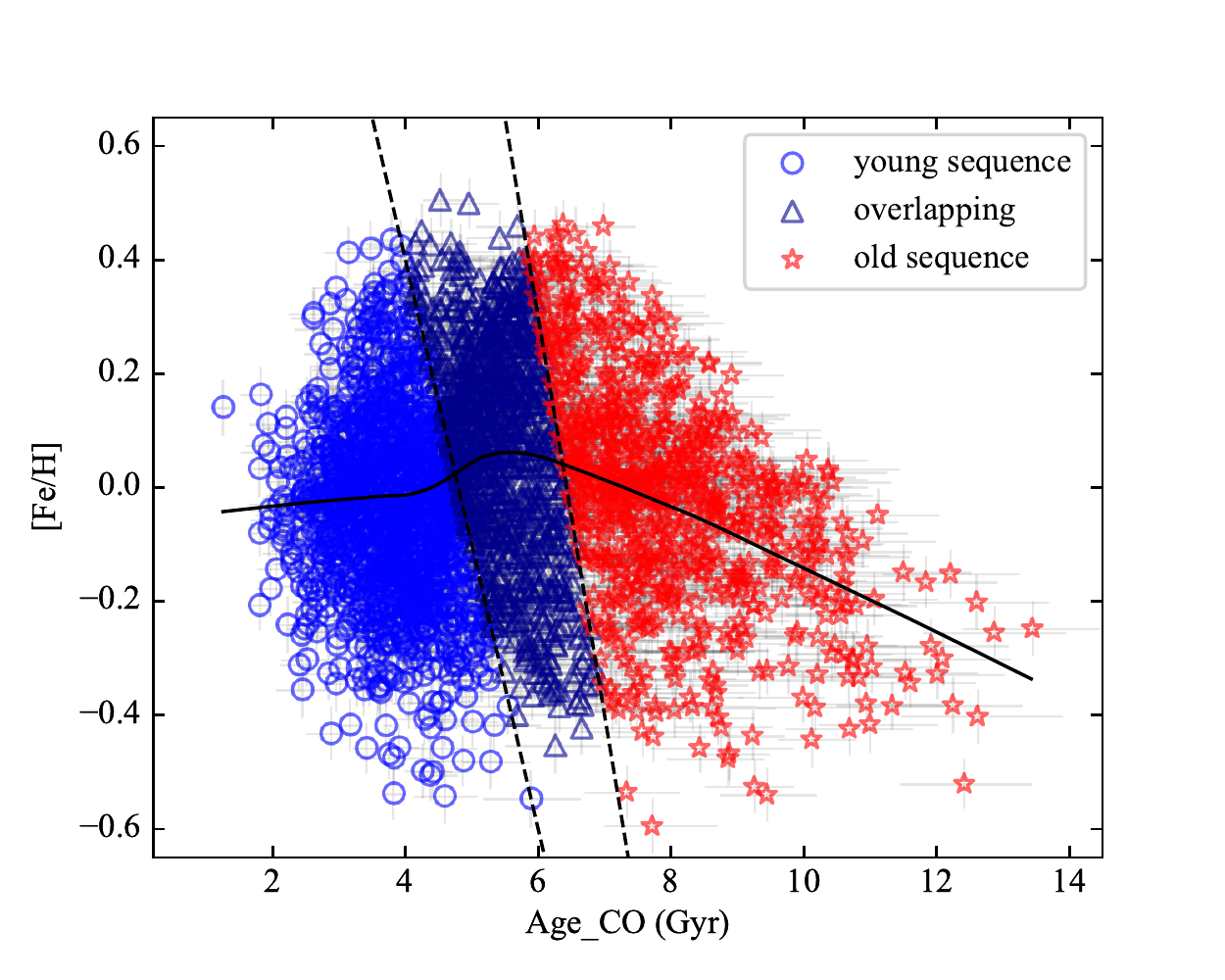}
\caption{[Fe/H] as a function of age. The black dashed lines show the criterion for dividing stars into two sequences: an old sequence shown in red asterisks and a young sequence shown in blue circles. The stars in the overlapping region are shown in dark blue triangles. The black solid line represents the best local non-parametric regression.}
\label{fig8}
\end{figure}

Different elements are released to the interstellar medium by stars with
different masses and therefore on different time scales. Thus, abundances to age ratios could provide information about the past history of star formation and gas accretion for the Milky Way. Here we present abundances to age ratios in disk population with ages calculated considering the variation of C and O abundances. Figure \ref{fig8} shows the [Fe/H]-age diagram. We find a predominantly flat trend at age~$\leq$~7 Gyr, and a decreasing trend at age~$\geq$~7 Gyr, indicating the possible existence of two different sequences in our sample which is consistent with result of Figure~\ref{fig5}c. \citet{Nissen2020} (their Fig.3) also found two sequences in [Fe/H]-age plane for solar-type stars. Based on the different trends, we divide sample stars into a young sequence of stars with age mostly $<$ 7 Gyr (blue circles), and a relatively older sequence of stars with age mostly $>$ 7 Gyr (red asterisks), overlapping at 5 Gyr $\leq$~age~$\leq$ 7 Gyr (dark-blue triangles).
\begin{figure*}
\centering
\includegraphics[width=0.9\textwidth]{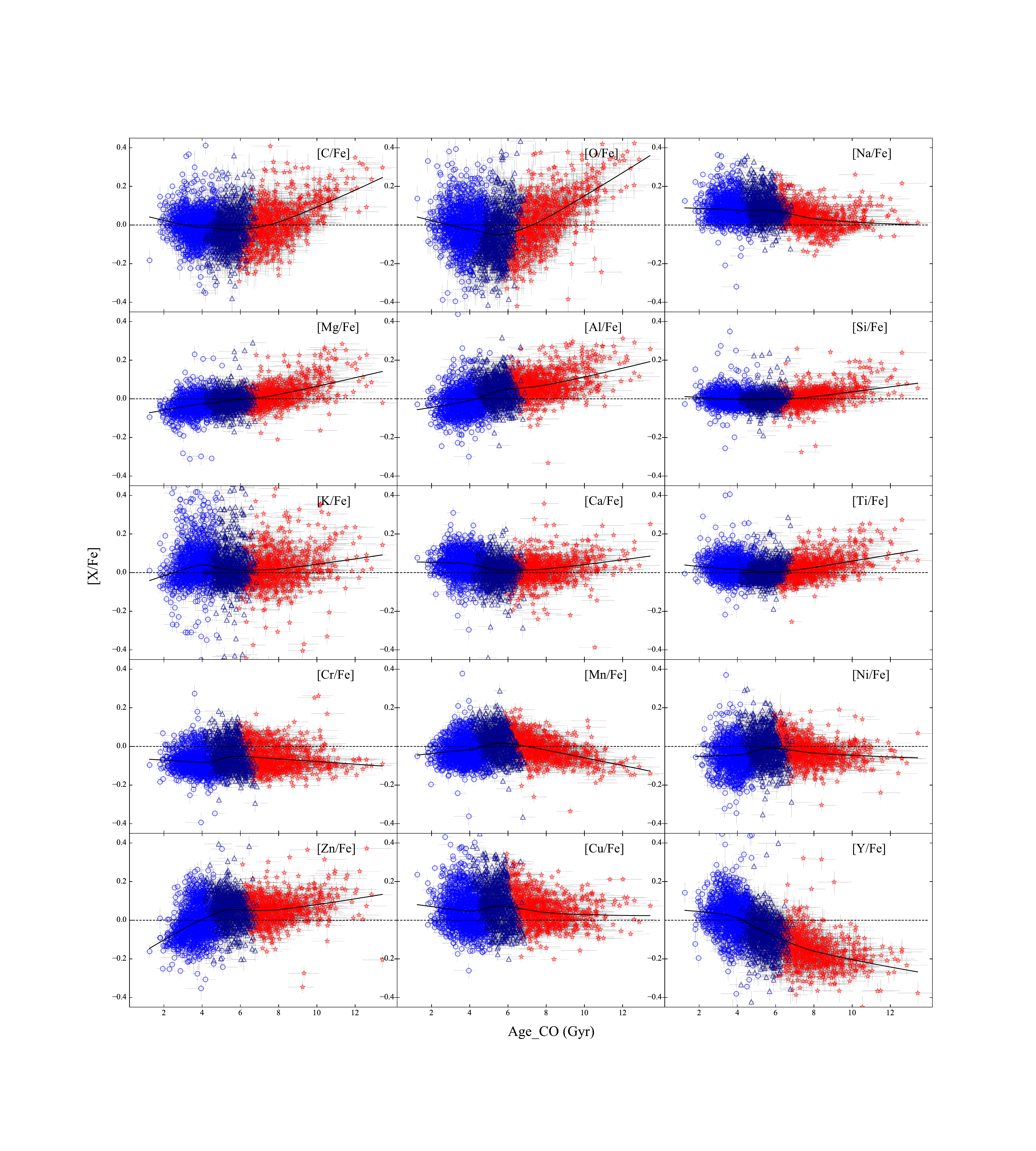}
\caption{Relations between various chemical-abundance ratios (indicated in each panel) and ages for our sample stars. The symbols are the same as those defined in Figure \ref{fig8}. The black solid line represents the best local non-parametric regression.}
\label{fig9}
\end{figure*}

Figure \ref{fig9} shows the relations between various chemical abundance ratios and age in the disk population. The trends of abundances to age ratios also show two sequences corresponding to the young sequence and the old sequence. The ratios of [C/Fe] and [O/Fe] slightly decrease with large scatter in the young sequence, and then start to increase with age in the old sequence. The tight correlation between [C/Fe] or [O/Fe] and age with small dispersion in the old sequence suggests that the ratios could be good age proxies for old stars. Recalling that the root cause(s) of the chemical evolution of C in the Galaxy is unclear (e.g. Type \uppercase\expandafter{\romannumeral2} SNe, stellar winds from massive stars such as Wolf–Rayet stars, intermediate-mass and low-mass stars in the planetary nebula phase, and stars at the end of the giant phase as mentioned in \citet{Nissen2013}) and the origin of O is exclusively by CCSNe \citep{Franchini2021}, the similarity of the two relations ([C/Fe]-age and [O/Fe]-age) could imply that C and O might be of similar sources. The Na, Al, K and Cu are mainly produced by exploding massive stars, but they show different trends in [X/Fe]-age plane. This is due to that they are odd Z elements which are strongly related to the metallicity of progenitors \citep{Kobayashi2020}. For $\alpha$ elements (Mg, Si, Ca and Ti), they are even-Z elements, mainly produced by exploding massive stars. The [Mg/Fe] shows increasing trends with small dispersion, indicating that [Mg/Fe] is a good age proxy for disk population. The [Si/Fe], [Ca/Fe], and [Ti/Fe] show flat trends in the young sequence, and increasing trends in the old sequence. For Cr and Zn, they are mainly produced by Type \uppercase\expandafter{\romannumeral1}a SNe, and they show the scattered distribution at all age. The Mn and Ni are iron-peak elements which are mainly produced by exploding massive stars and exploding white dwarf. They show increasing trends in the young sequence and decreasing trends in the old sequence. The [Y/Fe] shows a decline at all age, indicative of strong dependence on age. This is due to the delayed production from successive captures of neutrons by iron-peak elements in low-mass AGB stars with respect to the early contribution of SNe \uppercase\expandafter{\romannumeral1}a and SNe \uppercase\expandafter{\romannumeral2} that produce iron \citep{casali2020}.

\begin{figure}
\centering
\includegraphics[width=8cm]{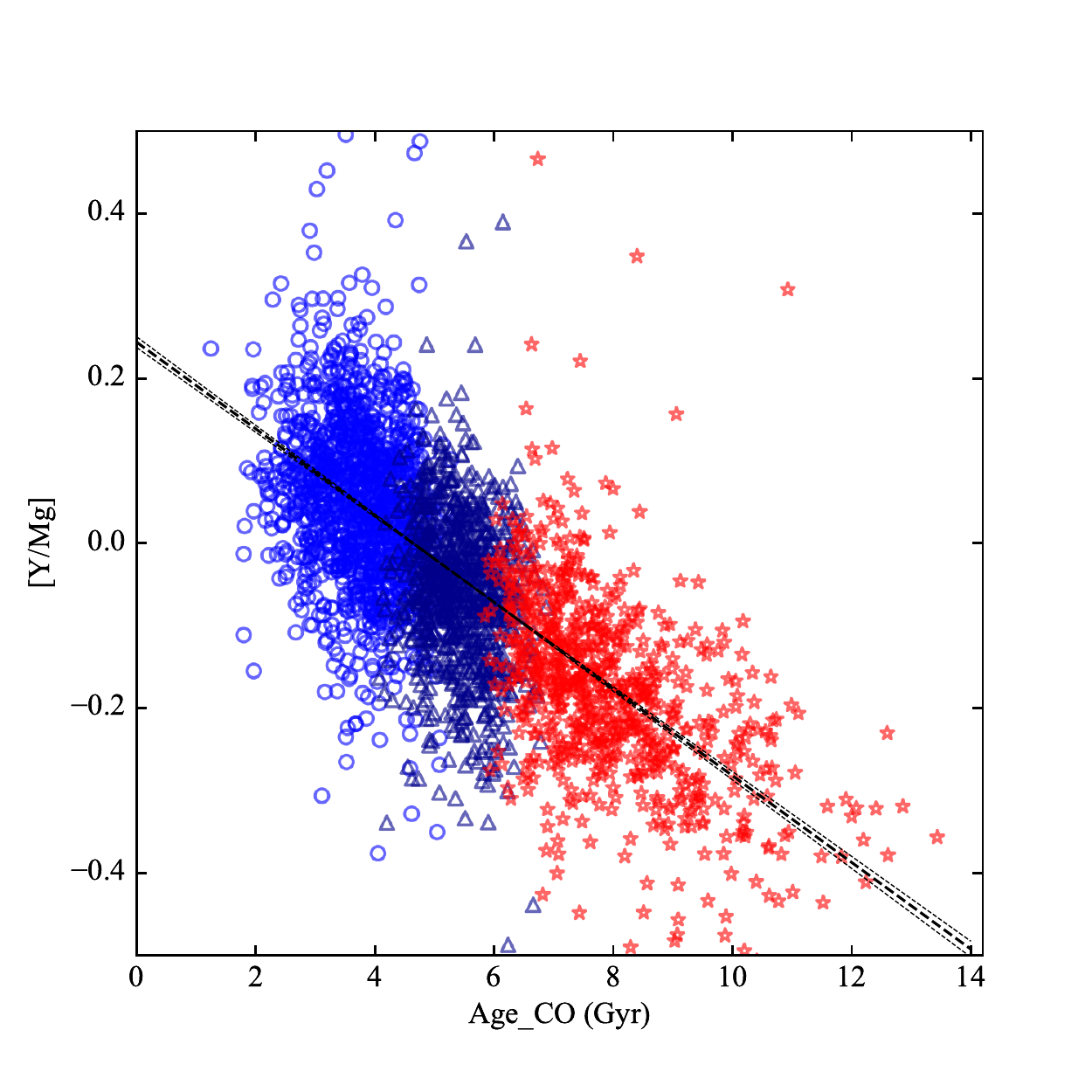}
\caption{[Y/Mg]-age relation and its best-fitting linear model (black dashed line) and 1-$\sigma$ region (between two dotted lines). The symbols are the same as those defined in Figure \ref{fig8}.}
\label{fig10}
\end{figure}
\begin{figure}
\centering
\includegraphics[width=8cm]{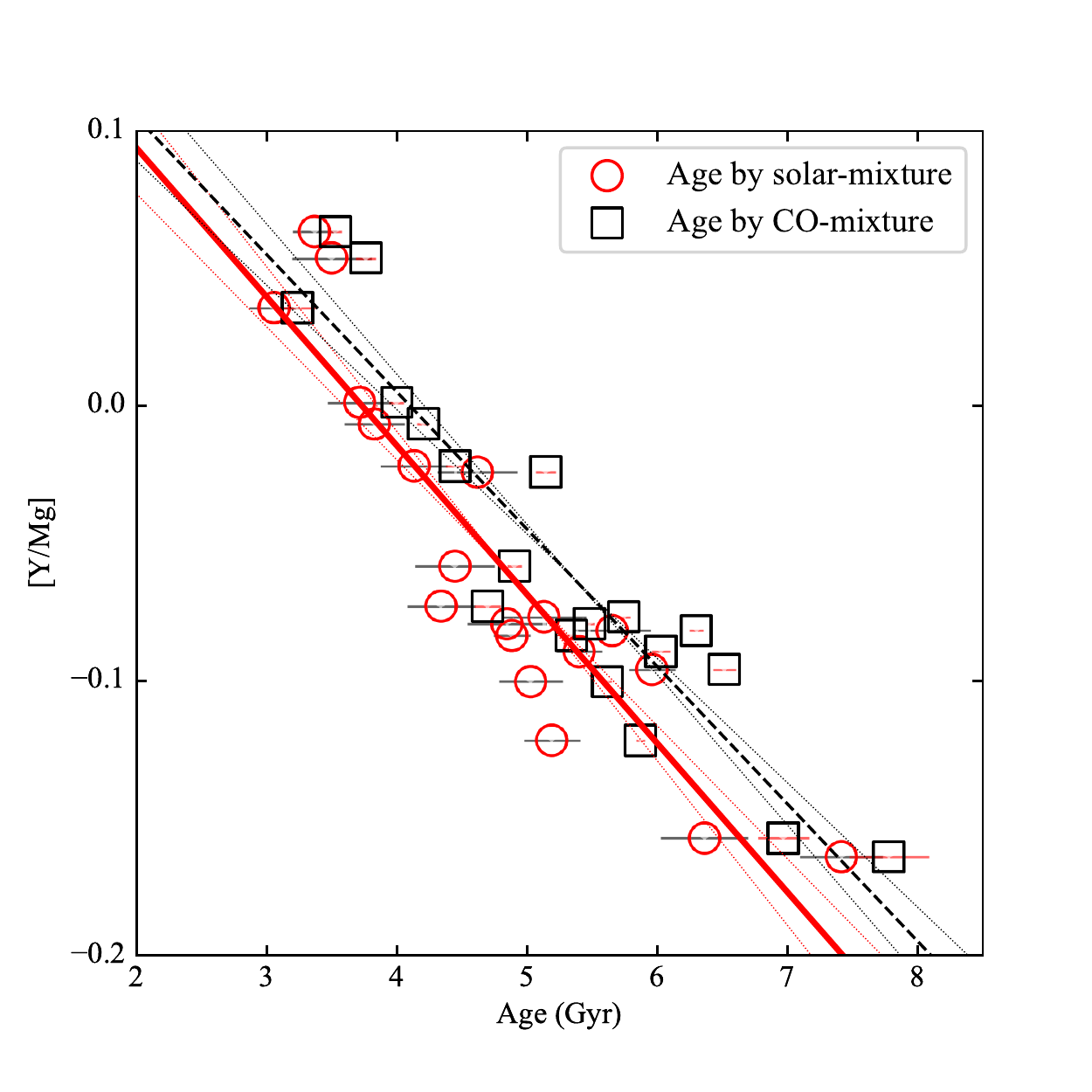}
\caption{[Y/Mg] as a function of age for CO-poor stars; the points indicate the mean age and the mean [Y/Mg] in each bin; the error bars indicate the standard deviation of each bin. Black squares: result by CO-mixture. Red circles: result by solar-mixture. Black dashed lines: linear fitting for the CO-mixture results. Red solid lines: linear fitting for result by solar-mixture. The black dotted lines represent the margin of error of the linear fit for the CO-mixture results. The red dotted lines represent the margin of error of the linear fit for result by solar-mixture.}
\label{fig11}
\end{figure}

\begin{table*}
  \tiny
  \centering
  \caption{Fundamental and Kinematic parameters for the whole sample determined in this work.}
  \label{tab6}
  \renewcommand\tabcolsep{4.0pt} %
  \begin{tabular}{ccccccc}
  \hline
  Star & $Mass$ & ${Age_{\rm CO}}$ & ${Age}$ & ${U_{\rm LSR}}$ & ${V_{\rm LSR}}$ & ${W_{\rm LSR}}$ \\

   ID & ($M_\odot$)  & (Gyr) & (Gyr) & (km/s)& (km/s) & (km/s) \\																																							
  \hline
  11285775-0440070 & $1.18_{-0.02}^{+0.04}$&$4.24_{-0.50}^{+0.56}$&$3.83_{-0.36}^{+0.65}$&-34.3&	-50.0	&	11.5\\
  17215393+0618245 & $1.20_{-0.02}^{+0.02}$&$3.21_{-0.66}^{+0.55}$&$2.99_{-0.65}^{+0.50}$&	-9.41	&	-52.8	& -5.39\\	
  $\vdots$&$\vdots$&$\vdots$&$\vdots$&$\vdots$&$\vdots$&$\vdots$\\
  \hline

  \end{tabular}
  {\\Note. This table is available in its entirety in machine-readable form.}
\end{table*}

\begin{table}
  \centering
  \caption{Spearman correlation coefficients, $\rho$, of [X/Fe] or [Y/X] abundance ratios vs. the stellar age.}
  \label{tab7}
  \begin{tabular}{cc}
  \hline
   Element & $\rho$  \\		
  \hline			
   $\rm{[Mg/Fe]}$ &0.455	\\			
   $\rm{[Al/Fe]}$ &0.523\\			
   $\rm{[Si/Fe]}$ &0.125	\\			
   $\rm{[Ca/Fe]}$ &-0.219	\\			
   $\rm{[Ti/Fe]}$ &0.012	\\			
   $\rm{[Y/Fe]}$ &-0.656	\\			
   $\rm{[Y/Mg]}$ &-0.701	\\			
   $\rm{[Y/Al]}$ &-0.694	\\			
   $\rm{[Y/Si]}$ &-0.639	\\			
   $\rm{[Y/Ca]}$ &-0.517	\\			
   $\rm{[Y/Ti]}$ &-0.606	\\			
  \hline
  \end{tabular}
\end{table}

The chemical clocks are empirical relations which can derive stellar ages from chemical abundances. Based on 1111 dwarfs stars from the HARPS-GTO program, \citet{Delgado2019} propose that any ratio of [Y or Sr] over $\alpha$ elements (plus Zn and Al) is a good candidate to be a chemical clock. Similarly as \citet{Delgado2019}, we use the Spearman correlation coeffcient ($\rho$) to study the correlation of the chemical species with age. Table \ref{tab7} shows the values of $\rho$ between different chemical species and age. Generally, the higher $|\rho|$ corresponds to the better linear correlation. For [Si/Fe], [Ca/Fe] and [Ti/Fe], their $|\rho|$ are extremely low, indicating that they are not chemical clocks. The [Y/Mg] has the highest $|\rho|$, corresponding to the strongest anti-correlation with age. Therefore [Y/Mg] can be a good chemical clock for disk population. Figure \ref{fig10} shows the relation of [Y/Mg] versus age. We present the polynomial fit on sample stars and the formula is:
\begin{equation}\label{eq9}
\rm{[Y/Mg]} = -0.053 (\pm 0.001)*{\rm{Age_{CO}}}+0.244 (\pm 0.006)
\end{equation}

The C and O can influence the age determination, therefore it is significant to study the effect by C an O on chemical clocks. We illustrate the age bins for CO-poor stars. The CO-poor stars are sorted by their age and then divided into 19 bins with each bin containing 20 stars (the last bin contains 24 stars). Figure \ref{fig11} shows the relation of [Y/Mg] versus age for CO-poor stars. We perform the polynomial fit for all age bins and the specific formula are:
\begin{equation}\label{eq10}
\rm{[Y/Mg]} = -0.054 (\pm 0.006)*{\rm{Age}}+0.202 (\pm 0.028)
\end{equation}
\begin{equation}\label{eq11}
\rm{[Y/Mg]} = -0.050 (\pm 0.005)*{\rm{Age_{CO}}}+0.205 (\pm 0.025)
\end{equation}
The age bins by CO-mixture are systematically older than those by solar-mixture, causing changes of slope and intercept in [Y/Mg]-age relation for CO-poor stars. The fractional age difference between two models can be simply given by:
\begin{equation}\label{eq12}
\frac{\rm{Age_{CO}}-\rm{Age}}{\rm{Age}}= \frac{3}{37}+\frac{0.057}{-18.5*{\rm{[Y/Mg]}}+3.74}
\end{equation}
For CO-poor stars, the C and O can influence the age determination based on chemical clocks globally by $\geq$~8\%.

\subsection{Kinematic Analysis: Spatial Velocity versus Age}
We determine kinematic properties of sample stars with the Gaia EDR3 database \citep{Gaia Collaboration2020}. For sample stars, the distance are derived from \citet{Bailer-Jones2021}. The proper motion are obtained from Gaia EDR3 database; the radial velocity is given by GALAH DR3 \citep{buder2021}. The space velocity components, $U_{LSR}$, $V_{LSR}$, $W_{LSR}$ are calculated with respect to the local standard of rest, adopting the standard solar motion   (U, V, W) = (-8.5, 13.38, 6.49) (km/s) \citep{Coskunoglu2011}. All the kinematic properties of our sample are listed in Table \ref{tab5}.

\begin{figure}
\centering
\includegraphics[width=8cm]{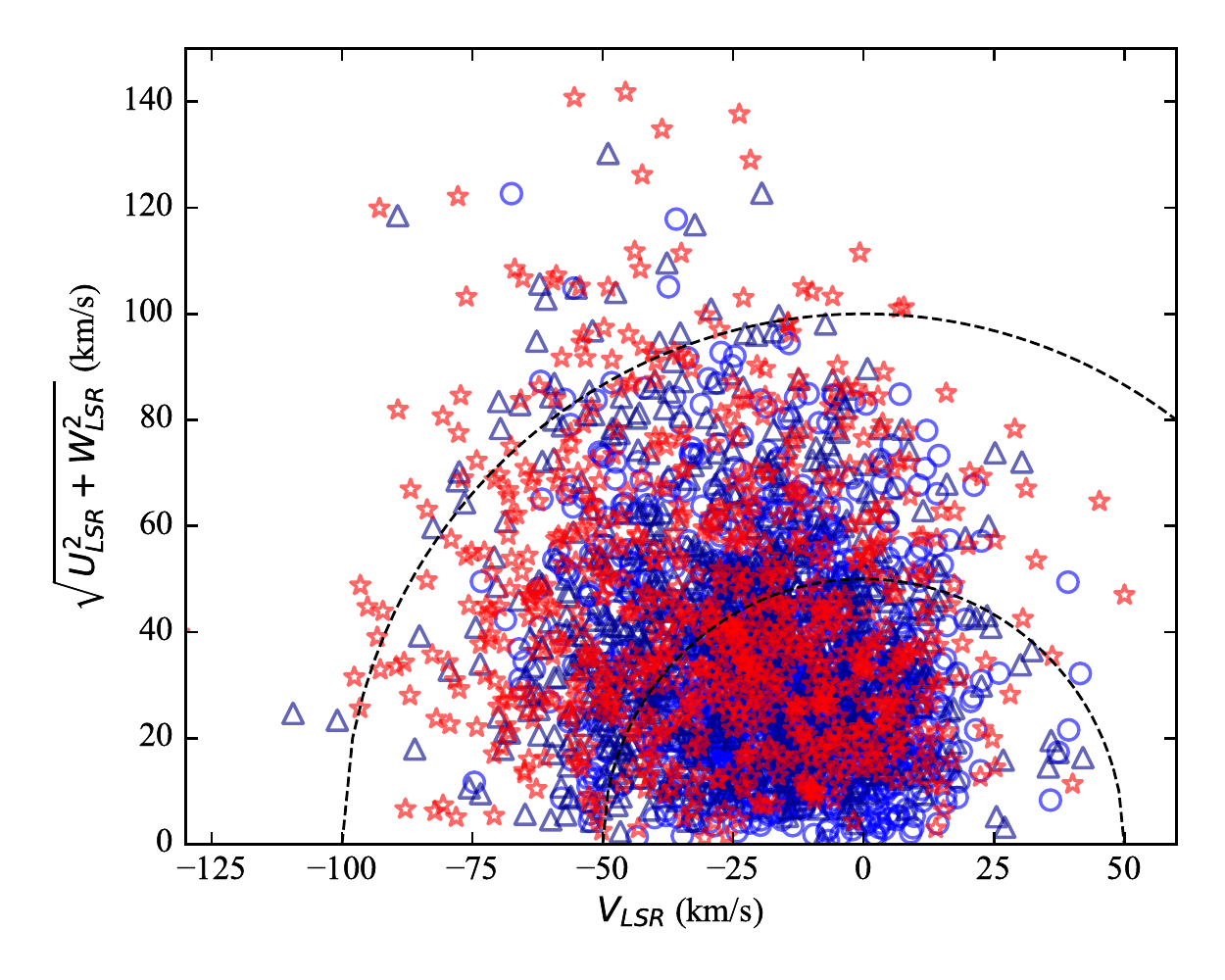}
\caption{Toomre diagram of our sample stars.  The dashed lines correspond to respectively $\rm{V_{total}}$ = 50 km/s and $\rm{V_{total}}$ = 100 km/s. The symbols are the same as those in Figure \ref{fig8}.}
\label{fig12}
\end{figure}

The Toomre diagram is widely used to divide thin-disk and thick-disk populations in kinematic space \citep[e.g.][]{Adibekyan2012,Buder2019}. Figure \ref{fig12} shows the Toomre diagram for sample stars. Most stars show solar-disk like motion because their U, V, and W are similar to the local standard of rest. Stars of the young sequence and the old sequence show similar behavior, indicating no clues of separating these two sequences based on kinematics. This should be due to the kinematic evolution of the Galaxy which altered the original spatial and kinematic distributions of stars \citep{Nissen2013}. As we mentioned in Section~\ref{sec2.1}, the young sequence and the old sequence can not be separated based on chemistry either, therefore stellar age is an important parameter to distinguish these two sequences in our sample.

\begin{figure}
\centering
\includegraphics[width=8cm]{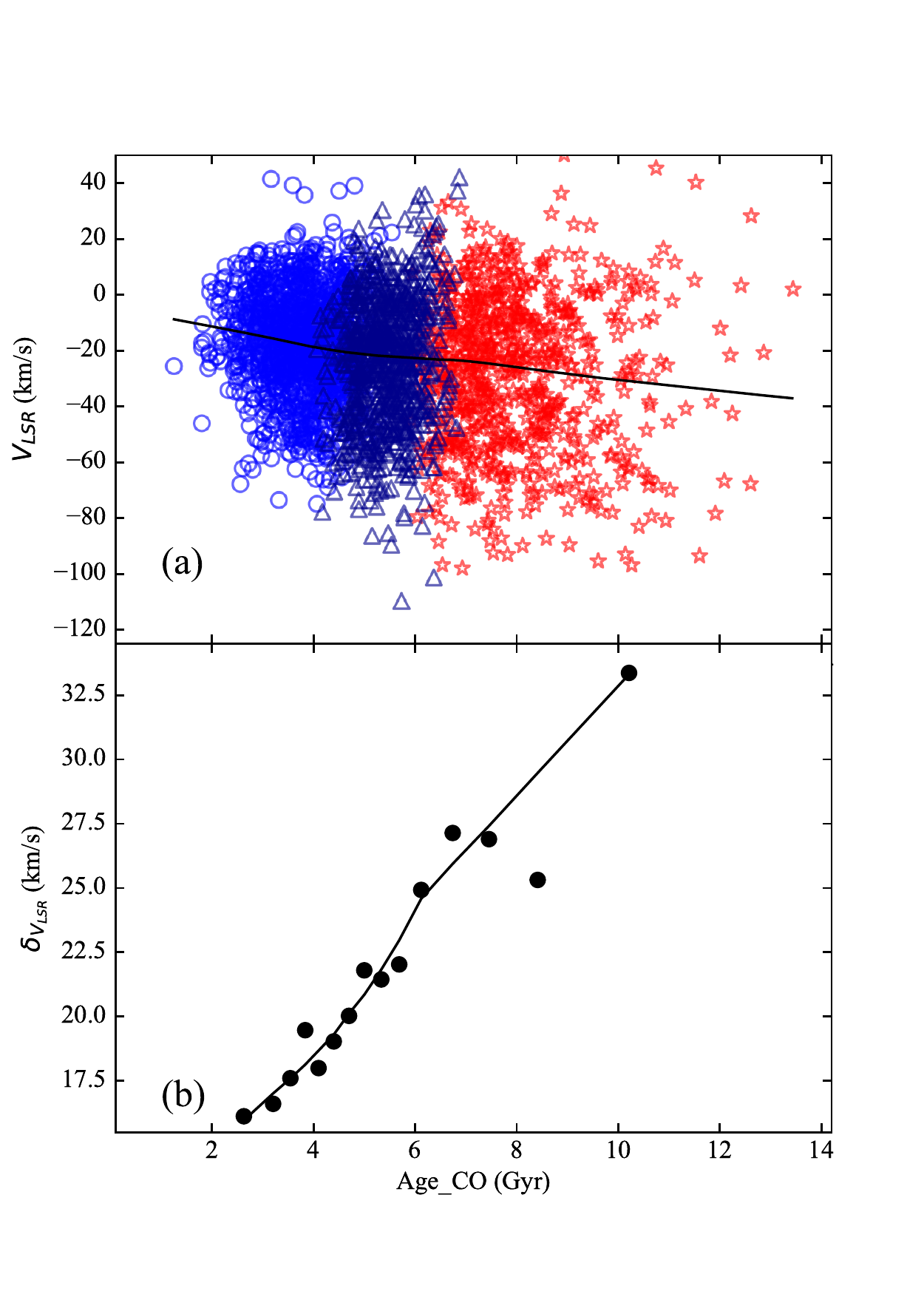}
\caption{\textbf{Top}: Line-of-sight velocity ($V_{LSR}$ ) as a function of age. The symbols are the same as those in Figure \ref{fig8}. \textbf{Bottom}: Standard deviation of binned $V_{LSR}$ as a function of binned age. The black solid lines represent the best local non-parametric regression. }
\label{fig13}
\end{figure}

Figure \ref{fig13} shows relations between $V_{LSR}$, the scatter of $V_{LSR}$ and age. Panel (a) of Figure \ref{fig13} shows that the $V_{LSR}$ becomes clearly more scattered in the old sequence compared to that in the young sequence. We also illustrate the age bins for sample stars. The stars are sorted by their age and then divided into 15 bins with each bin containing 200 stars (the last bin contains 126 stars). For each bin, we calculate the standard deviation of the $V_{LSR}$ ($\delta_{V_{LSR}}$). Panel (b) of Figure \ref{fig13} shows the relation of the scatter of $V_{LSR}$ versus age. The $\delta_{V_{LSR}}$ increases with age in disk population, which is also found in many previous works \citep[e.g.][]{Almeida-Fernandes2018}.

\section{Conclusion}\label{sec5}
The MSTO stars are good tracers of Galactic populations because their ages can be reliably obtained. Based on the GALAH survey, we select a sample of 2926 MSTO stars and determine their ages with mean age uncertainty of $\sim$~10\%. The age distribution of sample stars shows two different groups: a young group with mean age $\sim$4.5 Gyr and an old group with mean age $\sim$8 Gyr.

We estimate ages of 384 CO-poor stars and 18 high-O stars considering the variation of C and O abundances. Comparing result by $\alpha$-mixture or solar-mixture, ages of most high-O stars calculated by CO-mixture are younger. Ages of CO-poor stars are systematically affected by $\sim$10\%, globally shifting to $\sim$0.5 Gyr older compared to the results using solar metal-mixture. The age difference increases with [Fe/H] or the absolute value of [O/Fe]. Of the stars with [Fe/H]$\sim$0.3-0.5 or [O/Fe]~$\leq$~-0.25, many have age differences $\geq$~20\%, and even reach up to 36\%.

The [Fe/H]-age relation shows three different trends, indicating the possible existence of two distinct sequences. Based on the different trends, we divide sample stars into a young sequence of stars with age mostly $<$ 7 Gyr, and a relatively older sequence of stars with age mostly $>$ 7 Gyr, overlapping at 5 Gyr $\leq$~age~$\leq$ 7 Gyr. These two sequences also show different trends in [X/Fe]-age planes, especially in [O/Fe]-age plane. The [C/Fe] and [O/Fe] show similar behavior, indicating that they might be of similar sources. The [O/Fe] correlate with age in the old sequence, indicative of a good age proxy for old stars.

We use the Spearman correlation coeffcient to study the correlation of the chemical species with age. The [Y/Mg] has the strongest correlation with age, indicating that [Y/Mg] is a good chemical clock for disk population. We calculate the empirical relation between [Y/Mg] and stellar age in our sample, and the specific formula is: [Y/Mg] = -0.053 $(\pm 0.001)*{\rm{Age_{CO}}}+0.244 (\pm 0.006)$. For CO-poor stars, stellar ages based on chemical clocks can be affected by $\geq$8\% due to C and O abundances.

Based on Gaia EDR3 database, we calculate the space velocity components, $U_{LSR}$, $V_{LSR}$, $W_{LSR}$ with respect to the local standard of rest for our sample stars. Most stars show solar-like motion. The young sequence and the old sequence can not be separated based on chemistry or kinematics, and stellar age is an important parameter to distinguish these two sequences. In $V_{LSR}$-age plane, the old sequence shows clearly more scattered trend compared to that of the young sequence. The scatter of $V_{LSR}$ increases with age in disk population.

\nolinenumbers
\begin{acknowledgements}
This work is supported by the Joint Research Fund in Astronomy (U2031203,U1631236) under cooperative agreement between the National Natural Science Foundation of China (NSFC) and Chinese Academy of Sciences (CAS). This work is also supported by Nos. 12090040, 12090042, 11625313 and 11988101.

This work used the data from the GALAH survey which is based on observations made at the Anglo Australian Telescope, under programmes A/2013B/13, A/2014A/25, A/2015A/19, A/2017A/18.

This work has made use of data from the European Space Agency (ESA) mission Gaia (https://www.cosmos.esa.int/gaia), processed by the Gaia Data Processing and Analysis Consortium (DPAC, https://www.cosmos.esa.int/web/gaia/dpac/consortium).
\end{acknowledgements}

\end{document}